\documentclass[conference]{IEEEtran}
\IEEEoverridecommandlockouts
\usepackage{amsmath,graphicx,algorithm,algorithmic}

\usepackage{amsmath,amssymb,amsthm,amsfonts,latexsym,bbm,xspace,graphicx,float,mathtools,paralist,verbatim,xcolor}
\usepackage{url}
\PassOptionsToPackage{hyphens}{url}\usepackage{hyperref}

\newcommand{\prob}{\mathbb{P}}
\newcommand{\empprob}{\widehat{\prob}}

\newcommand{\probe}{\alpha}
\newcommand{\response}{\beta}
\newcommand{\utility}{u}

\newcommand{\reals}{\mathbb{R}}
\newcommand{\dataset}{\mathcal{D}}
\newcommand{\statenoisecov}{Q}
\newcommand{\obsnoisecov}{R}
\renewcommand{\time}{k}
\newcommand{\horizon}{K}
\newcommand{\sigdim}{m}

\newcommand{\probedim}{m}
\newcommand{\nresponse}{\hat{\response}}
\newcommand{\pertresponse}{\tilde{\response}}
\newcommand{\idresponse}{\response^\ast}
\newcommand{\ndataset}{\widehat{\dataset}}
\newcommand{\vecresponse}{\boldsymbol{\response}}
\newcommand{\pdf}{f}

\newcommand{\iter}{i}

\newcommand{\argmin}{\operatorname{argmin}}
\newcommand{\argmax}{\operatorname{argmax}}

\newtheorem{definition}{Definition}
\newtheorem{theorem}{Theorem}

\def\BibTeX{{\rm B\kern-.05em{\sc i\kern-.025em b}\kern-.08em
    T\kern-.1667em\lower.7ex\hbox{E}\kern-.125emX}}
\begin{document}

\title{Meta-Cognition. An Inverse-Inverse Reinforcement Learning Approach for Cognitive Radars
\thanks{This research was supported in part by a research contract from  Lockheed Martin, the Army Research Office grant W911NF-21-1-0093 and the Air Force Office of Scientific Research grant FA9550-22-1-0016.\\A short version of partial results appears in IEEE International Conference on Acoustics, Speech and Signal Processing (ICASSP), 2022.}
}

\author{\IEEEauthorblockN{Kunal Pattanayak}
\IEEEauthorblockA{\textit{Electrical and Computer Engineering} \\
\textit{Cornell University}\\
New York, US \\
kp487@cornell.edu}
\and
\IEEEauthorblockN{Vikram Krishnamurthy}
\IEEEauthorblockA{\textit{Electrical and Computer Engineering} \\
\textit{Cornell University}\\
New York, US \\
vikramk@cornell.edu}
\and
\IEEEauthorblockN{Christopher Berry}
\IEEEauthorblockA{
\textit{Lockheed Martin}\\\textit{Advanced Technology Laboratories}\\
New Jersey, US \\
christopher.m.berry@lmco.com}
}

\maketitle
\begin{abstract}
This paper considers meta-cognitive radars in an adversarial setting. A cognitive radar optimally adapts its waveform (response) in response to maneuvers (probes) of a possibly adversarial moving target.
A {\em meta-cognitive} radar is aware of the adversarial nature of the target and seeks to mitigate the adversarial target. {\em How should the meta-cognitive radar choose its responses to sufficiently confuse the adversary trying to estimate the radar's utility function?}
This paper abstracts the radar's meta-cognition problem in terms of the spectra (eigenvalues) of the state and observation noise covariance matrices, and embeds the algebraic Riccati equation into an economics-based utility maximization setup. This adversarial target is an {\em inverse reinforcement learner}. By observing a noisy sequence of radar's responses (waveforms), the adversarial target uses a statistical hypothesis test to detect if the radar is a utility maximizer.
In turn, the meta-cognitive radar deliberately chooses sub-optimal responses that increasing its Type-I error probability of the adversary's detector. We call this counter-adversarial step taken by the meta-cognitive radar as inverse inverse reinforcement learning (I-IRL).
We illustrate the meta-cognition results of this paper via simple numerical examples. Our approach for meta-cognition in this paper is based on
revealed preference theory in micro-economics
and inspired by results in differential privacy and adversarial obfuscation in machine learning.

\end{abstract}

\section{Introduction}
In abstract terms a meta-cognitive radar is a sophisticated constrained utility maximizer with multiple sets of utility functions and constraints that allow the radar to deploy different strategies depending on changing environments. Such radars adapt their waveform, scheduling and beam by optimizing  the utility functions in different situations. If a smart adversary can estimate the utility function of the radar, then it can exploit this information to mitigate the radar's performance (e.g., jam the radar with purposefully designed interference). A natural question is: how can the cognitive radar hide its utility function from the adversary by acting dumb?

\begin{figure*}
    \centering
    \includegraphics[width=0.9\linewidth]{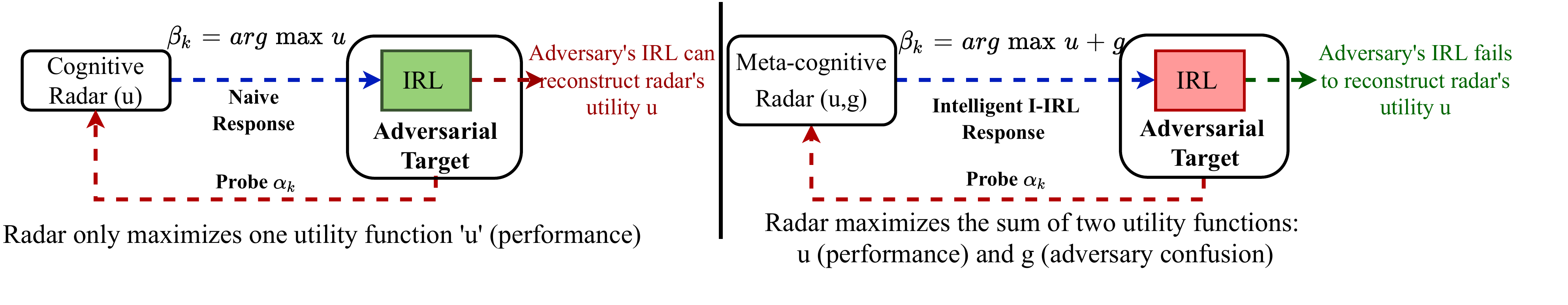}
    \caption{Schematic of the I-IRL approach for meta-cognition. The adversarial target sends a sequence of probe signals to the radar and records its responses. A cognitive radar (Left) chooses its optimal waveform that maximizes its utility function, but can be estimated by the adversary using IRL. The {\em meta-cognitive} radar deliberately chooses sub-optimal responses by trading off between maximizing its utility $u$ and confusing the adversary's IRL algorithm (abstracted by a second utility function $g$) (Right). This trade-off by the meta-cognitive radar ensures a poor reconstruction of the radar's performance utility function $u$ by the adversarial IRL algorithm; see Theorem~\ref{thrm:meta_result} for an informal statement of the I-IRL result.}
    \label{fig:schematic}
\end{figure*}

This paper investigates the interaction of a meta-cognitive radar and a smart adversary; see Fig.\,\ref{fig:schematic}. We formulate this interaction as an
{\em inverse-inverse reinforcement learning} problem.
Reinforcement learning (RL)~\cite{SUT18,KA18} deals with learning the optimal decision making strategy  by observing the response to a control input.
{\em Inverse}  reinforcement learning (IRL)~\cite{ABB01,Afr67,Var83} is the problem of reconstructing the utility function of a decision maker by observing its actions, namely, how can a smart adversary estimate the utility functions and constraints of a radar by observing its radiated pulses. Inverse IRL (I-IRL) is a natural extension of IRL: {\em If a radar knows that an
adversary is using an IRL algorithm to reconstruct the radar's  utility function by observing the radar's actions,  how can the radar deliberately distort its actions so that the adversary has a poor quality reconstruction of the radar's utility function?}\footnote{Though not discussed in this paper, an immediate extension is to formulate the radar-adversary interaction as a game, and is a topic of current research.}

\noindent {\bf Context and Related Work.} This paper comprises two interacting entities, a radar and an adversary. Each entity aims at mitigating the other and results in a sequence of strategies, which can be categorized as RL, IRL or I-IRL.
We discuss each aspect below in the context of cognitive radars and relate it to existing models for information fusion and information processing paradigms. \\
1. {\em Cognitive radar: RL agent.} A cognitive radar~\cite{Hay06,Hay12,BBSJR15}  uses the perception-action cycle of cognition to sense the environment and learn from it relevant information about the target and the environment. The cognitive radar is a reinforcement learner that maximizes its utility \cite{KAEM20,KPGKR21} and tunes its sensor to optimally satisfy its mission objectives. In the context of DFIG (Data Fusion Information Group) process model~\cite{DFIG1}, sensor adaptation by the radar can be viewed as Level 4-Process Refinement in the DFIG model.

\noindent 2. {\em Adversary: IRL agent for cognitive radar.} The IRL adversary observes the radar's responses and aims to estimate its utility function. \cite{KAEM20} propose revealed preference based IRL algorithms for identifying a cognitive radar performing optimal beam forming and optimal waveform adaptation. \cite{KR19} propose an IRL algorithm at the Bayesian tracker (Kalman filter) level using inverse filtering for estimating the system parameters of a cognitive radar. The IRL algorithm in this paper operates at a higher level of abstraction and estimates the radar's utility at the strategy level. \cite{PK20inverse} propose a Bayesian IRL algorithm for identifying the utility function of a sequential decision making radar. The adversary's IRL algorithm can be viewed as an electronic countermeasure (ECM)~\cite{BD78,SC86} in electronic warfare, where the aim is to mitigate the cognitive radar by estimating its utility function.

\noindent 3. {\em Meta-cognitive radar: I-IRL for adversarial IRL (this paper).}
Unlike a cognitive radar, a meta-cognitive radar is aware that an adversarial IRL algorithm is trying to mitigate is operations. So the radar adapts its cognition by deliberately transmitting sub-optimal responses to confuse the adversary.
Our recent paper \cite{PKB21}  presents  a cognition-masking scheme for a cognitive radar when the adversary has accurate measurements of the radar's response. This paper generalizes \cite{PKB21} to the case where the adversary's measurements are noisy, and the adversary uses an IRL detector to detect utility maximization.

Meta-cognition can be viewed as a sophisticated form of  electronic counter countermeasure (ECCM) to ECM in electronic warfare. \cite{ECCMsurvey} provides a comprehensive list of ECCM techniques. \cite{ECCM1,ECCM2} propose waveform adaptation schemes to counter barrage jamming. \cite{stealth1,stealth2,stealth3} exploit frequency diversity for radio stealth in multi-target and moving target tracking. However, meta-cognitive strategies involving deliberate performance loss to confuse the adversary's ECM have not been explored previously.

Finally, in the context of DFIG \cite{DFIG1} (see point 1 above), meta-cognition can be viewed as a response to a Level 5-User Refinement (IRL by the adversary). As a result, this counter-adversarial measure falls under Level 6-Mission Management in the DFIG model.

{\em Outline and Main Results.} Sec.\,\ref{sec:background} presents the key IRL results from revealed preference theory in micro-economics, Theorems~\ref{thrm:rp} and \ref{thrm:noise_rp}. While Theorem~\ref{thrm:rp} is standard in micro-economics literature, Theorem~\ref{thrm:noise_rp} is a generalization of Theorem~\ref{thrm:rp} for datasets corrupted with additive noise. Sec.\,\ref{sec:radar_example} motivates optimal waveform adaptation by a cognitive radar tracking an adversarial target as a utility maximization problem. Sec.\,\ref{sec:iirl} contains our main I-IRL results for meta-cognitive radars. Specifically, if an adversary uses the results of Sec.\,\ref{sec:background} to estimate the radar's utility function, Theorems~\ref{thrm:irp} and \ref{thrm:noise_irp} provide a counter-adversarial strategy used by the meta-cognitive radar to deliberately choose sub-optimal responses that obfuscate the adversary's IRL algorithm. The key idea is to maximize a sum of two utility functions for I-IRL, one utility function for performance guarantees, the other utility function for ensuring the adversary's confusion; see Theorem~\ref{thrm:meta_result} for an abstract statement of the I-IRL result.
Finally, we illustrate the I-IRL results of Sec.\,\ref{sec:iirl} via simple numerical examples in Sec.\,\ref{sec:numericalresults}.

\section{Background. Inverse Reinforcement Learning (IRL) for Estimating Utility Function} \label{sec:background}
Our objective is to devise a meta-cognitive radar strategy that confuses a smart adversary, where the adversary (target) uses  IRL to mitigate the radar. Keeping in mind our abstraction that a meta-cognitive radar is a constrained utility maximizer, the main result in this paper involving interaction of the meta-cognitive radar and smart adversary is summarized as follows.

\begin{theorem}[Informal. I-IRL for Meta-cognitive radar] Suppose a cognitive radar's response $\response_\time^\ast$ at time $\time$ maximizes its performance utility function $u$ subject to an adversary-imposed constraint $c_\time(\response)\leq 0$. Let $\operatorname{IRL}(\response_\time^\ast,c_\time)\leq 0$ denote the adversary's IRL algorithm for reconstructing the radar's utility  $u$.
Then, the I-IRL response $\pertresponse_\time$ of the meta-cognitive radar is given by:
\begin{equation}
    \response_\time = \argmax_{c_\time(\response)\leq 0} u(\response) + g(\operatorname{IRL}(c_k,\response_\time)),
\end{equation}
where the radar's obfuscation utility function $g(\cdot)$ is monotone and increasing and is responsible for confusing the adversary's IRL algorithm. \label{thrm:meta_result}
\end{theorem}

Although abstract, Theorem~\ref{thrm:meta_result} sets the stage for the key I-IRL results of this paper, Theorems~\ref{thrm:irp} and \ref{thrm:noise_irp} in Sec.\,\ref{sec:iirl}. The main idea for I-IRL for the meta-cognitive radar stays the same: deliberately choose smart sub-optimal responses to confuse the adversary's IRL algorithm while ensuring minimal performance loss.

The most general class of IRL algorithms that can be deployed by the smart adversary belong to the class of revealed preference methods  \cite{Afr67,Var83} developed in micro-economics\footnote{Traditional works in IRL~\cite{NG00,ZB08} assume {\em a priori} the existence of a utility function that rationalizes the decision of a decision maker. Revealed preference is more fundamental - it gives necessary and sufficient conditions  for the existence of a utility function and then constructs a set-valued estimate for the utility function.}.  This section reviews the key results in revealed preference theory, namely,  Theorems~\ref{thrm:rp} and \ref{thrm:noise_rp}. In the next section, we discuss how the interaction between the cognitive radar and the smart adversary can be embedded into the revealed preference framework formalized in this section.
\subsection{Deterministic IRL. Revealed Preferences, Afriat's Theorem} \label{sec:det_irl}
Revealed preference in microeconomics theory~\cite{Afr67,Var83} studies non-parametric detection of utility maximization behavior. A utility maximizing agent is defined as:
\begin{definition}[\cite{Afr67,Afr87}]\label{def:UM}
An agent is  a {\em utility maximizer} if for every probe $\probe_\time\in \reals_+^\probedim$, the  response $\response_\time \in \reals^\probedim_+$ satisfies
\begin{equation}
  \response_\time\in \argmax_{\{\response:\probe_\time
 '\response \leq 1\}}\utility(\response)
\label{eqn:utilitymaximization}
\end{equation}
where $\utility(\response)$ is a {\em monotone} utility function.
 \end{definition}

In economics, $\probe_\time$ is the price vector and  $\response_\time$ the consumption vector. Then $\probe_\time' \response \leq 1$ is a natural budget constraint\footnote{The budget constraint $\probe_\time' \response \leq 1$ is without loss of generality, and can be replaced by $\probe_\time' \response \leq c$ for any positive constant $c$.
}  for a consumer with 1 dollar. Given a dataset of price and consumption vectors, the aim in revealed preference is to determine if the consumer is a utility maximizer (rational) in the sense of (\ref{eqn:utilitymaximization}).


The key result in revealed preference is Afriat's theorem ~\cite{Afr67,Afr87,Var83, Die12,Var12} stated in Theorem~\ref{thrm:rp} below. A remarkable property of Afriat's theorem is that it gives testable conditions that are both necessary and sufficient for a time series of probes and responses to be consistent with utility maximization behavior~\eqref{eqn:utilitymaximization}.
\begin{theorem}[Afriat's Theorem~\cite{Afr67}: IRL for utility maximization (Definition~\ref{def:UM})] Given a sequence of probes and responses $\dataset=\{(\probe_\time,\response_\time), \time\in \{1,2,\dots,\horizon\}\}$,
 the following statements are equivalent:
	\begin{compactenum}
	\item There exists a monotone, continuous and concave utility function that satisfies \eqref{eqn:utilitymaximization}.
		\item \underline{\em Afriat's Test:} There exist positive reals $\{u_t,\lambda_t\}_{t=1}^\horizon$ such that the following inequalities are feasible.
			\begin{equation}
				u_s-u_t-\lambda_t \probe_t' (\response_s-\response_t) \leq 0 \; \forall t,s\in\{1,\dots,\horizon\}.\
				\label{eqn:AfriatFeasibilityTest}
			\end{equation}
			The monotone, concave utility function
			given by
			\begin{equation}
				\utility(\response) = \underset{t\in \{1,2,\dots,\horizon\}}{\operatorname{min}}\{u_t+\lambda_t \probe_t'(\response-\response_t)\}
				\label{eqn:estutility}
            \end{equation}
            constructed using $u_t$ and $\lambda_t$ \eqref{eqn:AfriatFeasibilityTest} rationalizes $\mathcal{D}$ \eqref{eqn:utilitymaximization}.
          \item The data set $\mathcal{D}$ satisfies the Generalized Axiom of Revealed Preference (GARP), namely, for any $t\in\{1,2,\ldots,\horizon\}$, the following implication holds:
          \begin{equation}\label{eqn:GARP}
           \probe_t' \response_t \geq \probe_t' \response_{t+1} \quad \forall t\leq k-1 \implies \probe_k'  \response_k \leq \probe_k'  \response_{1}.
          \end{equation}
	\end{compactenum}
\label{thrm:rp}
\end{theorem}

Afriat's theorem tests for economics-based rationality based on a finite dataset of an agent's input-output response.
GARP in statement 4 of Theorem~\ref{thrm:rp} is highly signification in micro-economic theory and is stated here for completeness.
The feasibility of the set of inequalities (\ref{eqn:AfriatFeasibilityTest}) can be checked using a linear programming solver; alternatively GARP can be checked  using Warshall's algorithm with $O(\horizon^3)$ computations~\cite{Var06,Var82}. The reconstructed utility in~\eqref{eqn:estutility}  is not unique but a set-valued estimate since any monotone transformation of~\eqref{eqn:estutility} also satisfies Afriat's Theorem; that is, the utility function constructed is ordinal.

\subsection{Stochastic IRL. Afriat's Theorem for Noisy Responses} \label{sec:stoch_irl}
Afriat's Theorem~(Theorem~\ref{thrm:rp}) assumes the agent's actions are accurately measured by the inverse learner. In this section, motivated by the interaction of the radar and adversary, we relax this  assumption. We  assume the inverse learner's (adversary) measurements of agent (radar) actions are noisy. Specifically, we assume an additive noise model, where the measured response $\nresponse_\time$ at time $\time$ is related to the true response  $\response_\time$ as:
\begin{equation}
    \nresponse_\time = \response_\time + w_\time,~w_\time\sim \pdf_{w}, \label{eqn:noise_response}
\end{equation}
where $w_\time$ is an independent and identically distributed random variable with a known pdf $\pdf_w$ for all $\time=1,2,\ldots,\horizon$.
{\em How to generalize the IRL result of Theorem~\ref{thrm:rp} to the noisy case?} Our key result is Theorem~\ref{thrm:noise_rp} below
that outlines a statistical hypothesis test for detecting utility maximization behavior given noisy measurements of the agent's actions.

For our hypothesis test below, let $H_0$ and $H_1$ denote the null and alternate hypotheses that the noise-less dataset aggregated from the agent's actions satisfy, and not satisfy, respectively, the conditions~\eqref{eqn:AfriatFeasibilityTest} for utility maximization behavior. The two types of error that arise in hypothesis testing are:
\begin{align}
    &\textbf{Type-I error:}~\text{Reject } H_0 \text{ when }H_0\text{ is true}.\nonumber\\
    &\textbf{Type-II error:}~\text{Fail to reject } H_0 \text{ when }H_1\text{ is not true}. \label{eqn:hyptest_errors}
\end{align}

We now state Theorem~\ref{thrm:noise_rp}. The key feature is that the statistical test of Theorem~\ref{thrm:noise_rp} is parametrized by a tunable scalar $\alpha$ that upper bounds the test's Type-I error probability.

\begin{theorem}[Afriat's Theorem for Noisy Observations] Suppose an adversary observes a noisy response sequence $\{\nresponse_\time\}$ \eqref{eqn:noise_response} of the radar in response to its probe signals $\{\probe_\time\}$. Let $\ndataset$ denote the noisy dataset $\{\probe_\time,\nresponse_\time\}$.
Then, the Type-I error probability of the statistical hypothesis test below parametrized by scalar $\alpha\in[0,1]$ is upper bounded by $\alpha$.
\begin{align}
    &\boxed{F_L(\phi^*(\ndataset)) \lessgtr_{H_1}^{H_0} 1-\alpha},
    \label{eqn:stat_test}\\
    \text{where } & F_L(\cdot) : \text{cdf of r.v.\ L}= \max_{i,j} \probe_i'(w_i-w_j),\nonumber\\
    & \phi^{\ast}(\ndataset) = \min \epsilon,~u_s - u_t - \lambda_t \probe_t' (\nresponse_s - \nresponse_t) \leq \lambda_t \epsilon,\label{eqn:def_suffstat}\\
    & s,t\in\{1,2,\ldots,\horizon\},~u_s,u_t,\lambda_t > 0.\nonumber
\end{align}
\label{thrm:noise_rp}
\end{theorem}
The proof of Theorem~\ref{thrm:noise_rp} is omitted for brevity and can be found in \cite[Appendix]{KH12}. The key idea in the proof is to show that the random variable $L$ in \eqref{eqn:stat_test} upper bounds the sufficient statistic $\phi^\ast(\cdot)$~\eqref{eqn:stat_test}.
In Theorem~\ref{thrm:noise_rp}, the scalar $\alpha$ is the ``significance level'' of the statistical test \eqref{eqn:stat_test}. The sufficient statistic $\phi^\ast(\cdot)$ in \eqref{eqn:stat_test} is the minimum perturbation so that $\ndataset$ passes the Afriat's test~\eqref{eqn:AfriatFeasibilityTest} of Theorem~\ref{thrm:rp}.
The constrained optimization problem \eqref{eqn:def_suffstat} is non-convex since the RHS of the constraint is bilinear in the feasible variable. However, since the objective function depends only on a scalar, a 1-dimensional line search algorithm can be used to solve \eqref{eqn:def_suffstat}. That is, for any fixed value of $\epsilon$, the constraints in \eqref{eqn:def_suffstat} specialize to a set of linear inequalities for which feasibility is straightforward to check.

\section{Optimal Waveform Adaption: Cognitive Radars and Utility Maximization}
\label{sec:radar_example}

With the above background on revealed preferences, we are now ready to define the radar adversary interaction.
Our  working assumption is that the cognitive radar is a constrained utility maximizer - that is, it satisfies economics based rationality. This section reconciles the radar's rationality with a vital aspect of cognitive radars, namely, optimal waveform adaptation in response to the maneuvers of the adversarial target.
We abstract optimal waveform adaptation of a cognitive radar using a Kalman filter for target tracking into the utility maximization setup of Definition~\ref{def:UM}. Specifically, we will express the linear budget constraint of Definition~\ref{def:UM} in terms of the eigenvalues (spectra) of the state and observation noise covariances of the radar's state space model.

Linear Gaussian dynamics for a target’s kinematics~\cite{LJ03} and linear Gaussian measurements at the radar are widely assumed as a useful approximation~\cite{BLK08}. Hence, consider the following state space model for the radar:
\begin{equation}\begin{split}
x_{n+1} &= A x_n + w_n(\probe_\time), \quad x_0 \sim \pi_0
\label{eqn:kalman-sys}\\
y_n &= C x_n + v_n(\response_\time) ,
\end{split}\end{equation}
where $x_n \in \mathcal{X} = \mathbb{R}^X$ is the target state with initial density  $\pi_0 \sim \mathcal{N}(\hat{x}_0, \Sigma_0)$,
$y_n \in \mathcal{Y} = \mathbb{R}^Y$ is the radar's observation,
$w_n \sim \mathcal{N}(0, Q(\probe_\time))$ and $v_n \sim \mathcal{N}(0, R(\response_\time))$ are mutually independent, Gaussian noise processes.

The state noise covariance $Q$ is parameterized by the adversarial target's probe $\probe_\time$ and the observation noise covariance $R$ is parameterized by the radar's response $\response_\time$; \cite[Sec.\,III-B]{KAEM20} elaborates on the relation between radar's waveform and observation noise covariance $R$.
Note that the state-space equations \eqref{eqn:kalman-sys} consist of two subscripts $n,k$. The subscript $n$ indicates system updates at the tracker level (faster timescale), and the subscript $\time$ indicates the epoch (slower timescale) for the probe and response. When state $x_n$ represents the position and velocity in Euclidean space, $A$ is a block diagonal constant velocity matrix \cite{BP99}.
The state noise covariance $Q(\probe)$ in \eqref{eqn:kalman-sys} models acceleration maneuvers of the target parameterized by the probes $\probe$.

The radar estimates the target state $\hat{x}_n$ with covariance $\Sigma_n$ from observations $y_{1:n}$.
The posterior $\pi_n$ is updated recursively via the classical Kalman filter equations:
\begin{align*}
\hspace{-0.5cm}\Sigma_{n+1|n} &= A \Sigma_n A' + Q(\probe_\time),~K_{n+1} = C \Sigma_{n+1|n} C' + R(\response_\time)\\
\hspace{-0.5cm}\psi_{n+1} &= \Sigma_{n+1|n} C' K^{-1}_{n+1}, ~\hat{x}_{n+1} = A \hat{x}_n + \psi_{n+1} (y_{n+1} - C A \hat{x}_n)\\
\hspace{-0.5cm}\Sigma_{n+1} &= (I - \psi_{n+1} C)\Sigma_{n+1|n}.
\end{align*}
Assuming the model parameters \eqref{eqn:kalman-sys} satisfy the conditions that $[A, C]$ is detectable and $[A, \sqrt{Q}]$ is stabilizable,
the steady-state predicted covariance $\Sigma_\infty$ is the unique positive semi-definite solution of the \textit{algebraic Riccati equation} (ARE):
\begin{align}
\mathcal{A}(\probe_\time, \response_\time, \Sigma) = & 
-\Sigma + A(\Sigma -
 \Sigma C' [C \Sigma C' + R(\response)]^{-1} C \Sigma) A' \nonumber\\
 & \quad \quad + Q(\probe) = 0.\label{eqn:ARE}
\end{align}
Denote $\Sigma^*(\probe_\time, \response_\time)$ as the solution of the ARE given probe $\probe_\time$ and response $\response_\time$ at time $k$.

We assume that the radar maximizes a utility function $u$ to choose its optimal waveform at the start of every epoch $\time$. Now, suppose:
\begin{compactitem}
    \item the target probe $\probe$ is the vector of eigenvalues of the positive definite matrix $Q$
    \item the radar response $\response$ is the vector of eigenvalues of the positive definite matrix $R^{-1}$.
\end{compactitem}

$\response_\time(i)$ can be viewed as the measurement precision (amount of directed energy) of the radar in the $i^{\text{th}}$ mode. Similarly, $\probe_\time(i)$ can be viewed as the radar's incentive for considering the $i^{\text{th}}$ mode of the target. Put together, $\probe_\time'\response_\time$ measures the signal-to-noise ratio (SNR) of the radar. Thus, $\probe_\time'\response_\time\leq 1$ is effectively a bound on the radar's SNR. Hence, wrt the constrained utility maximization setup of Definition~\ref{def:UM}, the radar chooses the most precise observation noise covariance $R(\response_\time)$ such that its SNR lies below a particular threshold \footnote{see \cite{KAEM20} for a more detailed discussion on the linear budget in terms of the solution to the ARE \eqref{eqn:ARE}.}.

In summary, this section embeds the cognitive radar's functionality of optimal waveform adaptation into the constrained utility maximization setup of Definition~\ref{thrm:rp}. As a result, the adversarial target can use Theorems~\ref{thrm:rp} and \ref{thrm:noise_rp} for IRL and reconstruct the radar's utility function. In the next section, we state our key I-IRL result for a {\em meta-cognitive} radar. That is, how should the radar tweak its response so that the adversary's IRL algorithm is sufficiently confused, and hence, the radar's utility function cannot be estimated.
The key idea is for the radar to deliberately choose sub-optimal waveforms that optimally trades off between
ensuring the radar's utility function does not pass Afriat's test for utility maximization and minimizing the radar's utility loss due to deliberately chosen sub-optimal responses.

\section{Inverse IRL (I-IRL) for Meta-Cognitive Radars}
\label{sec:iirl}
This section presents our key meta-cognition results of this paper, namely, Theorems~\ref{thrm:irp} and \ref{thrm:noise_irp}. If the adversarial target uses IRL (Theorems~\ref{thrm:rp}, \ref{thrm:noise_rp}) to detect the radar's cognition, the meta-cognitive radar can deploy the I-IRL counter-adversarial measures of Theorems~\ref{thrm:irp} and \ref{thrm:noise_irp} to foil the adversarial actions.
Theorem~\ref{thrm:irp} achieves I-IRL in the noise-less setting when the adversary uses Theorem~\ref{thrm:rp} (Afriat's Theorem) for IRL. When the adversary receives noisy measurements of the agent and uses Theorem~\ref{thrm:noise_rp} for IRL, the agent achieves I-IRL via Theorem~\ref{thrm:noise_irp}.

\subsection{I-IRL for Afriat's Theorem (Theorem~\ref{thrm:rp})} \label{sec:det_iirl}
Our first meta-cognition result, namely, Theorem~\ref{thrm:irp} below achieves I-IRL when the adversary has accurate measurement of the radar's waveform response. The key idea is for the radar to deliberately transmit sub-optimal waveforms (in the sense of \eqref{eqn:utilitymaximization}) so that the radar's utility function passes the Afriat's inequalities~\eqref{eqn:AfriatFeasibilityTest} with a small margin, where the margin is a pre-specified parameter chosen by the radar. This way the true utility function of the radar is a low-confidence estimate of the adversary, and hence is effectively masked by the meta-cognitive radar by deliberately compromising on its performance.

\begin{theorem}[Meta-cognition for Afriat's Theorem (Theorem~\ref{thrm:rp})]
Suppose the radar optimizes a monotone, continuous utility function $u$, and the adversary uses Theorem~\ref{thrm:rp} to estimate the radar's utility function. Given the adversary's probe sequence $\{\probe_\time\}_{\time=1}^{\horizon}$, the radar's response sequence $\{\pertresponse_\time\}_{\time=1}^{\horizon}$ that masks its utility function is given by: \vspace{-0.3cm}
\begin{align}
 \pertresponse_{1:\horizon} &= \argmin_{\{\response_{1:\horizon}\in\reals^\probedim_+\}} \sum_{\time=1}^\horizon \|\response_\time - \response_\time^\ast\|_2^2, \label{eqn:irp}\\     u(\response_s) & \geq u(\response_t)  
 - \nabla_{\response} u(\response_t^\ast)'(\response_s -\response_t) + \epsilon,~\forall~s,t,\label{eqn:constraint_lowmargin}\\
\response_t &  \geq 0,~\probe_\time'(\response^\ast_\time - \response_\time)=0,~\forall~t. \label{eqn:constaint_nonnegative}
\end{align}
In \eqref{eqn:irp}, $\response_\time^\ast=\argmax_{\{\response\in\mathbb{R}^m_+:\probe_\time'\response\leq 1\}} u(\response)$ is the naive response that maximizes utility $u$ given probe signal $\probe_\time$.
The variable $\epsilon$ denotes the margin with which the radar's response passes Afriat's test for utility maximization.
\label{thrm:irp}
\end{theorem}

Theorem~\ref{thrm:irp} for meta-cognition in radars masks the radar's utility function by deliberately perturbing its responses so that the responses almost fail the Afriat's test for utility maximization (Theorem~\ref{thrm:rp}). The constraint \eqref{eqn:constraint_lowmargin} ensures the margin with which the radar's utility function $u$ passes Afriat's test is less than $\epsilon$. The constraint  \eqref{eqn:constaint_nonnegative} ensures the perturbed response $\pertresponse_\time$ does not violate the budget constraint $\probe_\time'\response\leq 1$.

A naive response sequence  ($\pertresponse_\time=\idresponse_\time,~\forall\time$ in \eqref{eqn:irp}) causes the radar's utility function to pass the Afriat's test \eqref{eqn:AfriatFeasibilityTest} by a large margin and is thus a high-confidence utility estimate for the adversary. Due to the meta-cognition scheme of Theorem~\ref{thrm:irp}, the radar's utility function passes the Afriat's test by a very small margin
, and is no more a high-confidence utility estimate for the adversary. \vspace{0.15cm}

\noindent {\em Degree of meta-cognition $\epsilon$ for Theorem~\ref{thrm:irp}.}
A smaller value of $\epsilon$ implies better cognition masking and higher performance degradation of the agent, and hence a higher degree of meta-cognition.
One extreme case is setting $\epsilon=0$. This results in maximal masking of the radar's utility function. That is, Afriat's inequalities~\eqref{eqn:AfriatFeasibilityTest} are infeasible, and hence, the radar is classified as non-cognitive by the adversarial target. However, this complete masking requires the radar to deviate maximally from its optimal behavior~$\{\idresponse_\time\}$. On the other extreme, setting $\epsilon\geq\epsilon_{\text{max}} = \min_{s,t}  u(\beta_t) + \nabla_{\beta}u(\beta_t)'(\beta_s-\beta_t) - u(\beta_s)$ requires zero perturbation in the radar's response, but also results in zero masking of the radar's utility function. $\epsilon_{\text{max}}$ is simply the margin with which the true utility function and adversary's probe sequence pass the Afriat's test.

\subsection{I-IRL for Noisy Afriat's Theorem (Theorem~\ref{thrm:noise_rp})}
\label{sec:stoch_iirl}
In this section, we present our second key result, I-IRL when the adversary observes the radar's responses in noise. Recall from Sec.\,\ref{sec:stoch_irl} that the adversary uses a statistical hypothesis test with a bounded Type-I error to detect if the radar is cognitive (utility maximizer) or not. Achieving I-IRL for meta-cognitive radars in a noisy setting generalizes Theorem~\ref{thrm:irp} where we assume zero measurement error in the radar's response measured by the adversary.
Intuitively, since the radar's responses are now measured in noise by the adversary, the radar can at best control the probability of the adversary detecting it as a utility maximizer, namely, the conditional Type-I error probability defined as:
\begin{align}
   & \prob(H_1 | \{\probe_\time,\response_\time,u\})
=  \prob(\phi_u^\ast(\{\probe_\time,\nresponse_\time\}) \geq F_L^{-1}(1-\alpha)),\text{ where}\nonumber\\
    &\phi^\ast_u(\{\probe_\time,\nresponse_\time\}) = \max_{s,t} (u(\nresponse_t)-u(\nresponse_s) + \lambda_t \probe_t'(\nresponse_s-\nresponse_t))/\lambda_t,\nonumber\\
&\nabla_{\response}u(\response_t^\ast) = \lambda_t \probe_t,~\nresponse_\time = \response_\time + w_\time,~w_\time\sim f_w.\label{eqn:surr_error}
\end{align}
In \eqref{eqn:surr_error}, $\probe_\time$ is the adversary's probe,  $\response_\time^\ast$ is the radar's naive waveform response that maximizes its utility $u$ and $\response_\time$ is the radar's transmitted I-IRL waveform that confuses the adversary.  The sufficient statistic in \eqref{eqn:surr_error} 
is not obtained by solving an optimization problem as in \eqref{eqn:def_suffstat}. This is due to the conditioning of the Type-I error probability on the utility function $u$ which implicitly sets the feasible variables $u_\time,\lambda_\time$ in \eqref{eqn:def_suffstat} to a fixed value in \eqref{eqn:surr_error}. Intuitively, \eqref{eqn:surr_error} can be viewed as the probability with which the radar's true utility function fails the Afriat's inequalities and serves as the degree of confusion of the adversary in our I-IRL result below.

We are now ready to state Theorem~\ref{thrm:noise_irp}. The meta-cognitive radar in Theorem~\ref{thrm:noise_irp} trades off between minimizing its utility loss due to sub-optimal choice of waveform (quality-of-service) in response to the adversary's probes and maximizing the conditional Type-I error probability~\eqref{eqn:surr_error} of the adversary's detector (degree of confusion).

\begin{theorem}[Meta-cognition for noisy Afriat's theorem (Theorem~\ref{thrm:noise_rp})] \label{thrm:noise_irp} Given the adversary's probe sequence $\{\probe_\time\}_{\time=1}^\horizon$. Let $\{\idresponse_\time\}_{\time=1}^{\horizon}$ denote the naive response sequence that maximizes the radar's utility $u$~(Definition~\ref{def:UM}). Suppose the adversary uses Theorem~\ref{thrm:noise_rp} for IRL. Then, the radar's I-IRL response sequence $\{\pertresponse_\time\}$ to confuse the adversary is given by:
\begin{align}\hspace{-0.5cm}
   \{\pertresponse_\time\} = & \min_{\{\response_\time:\probe_\time'\response_\time\leq 1\}} ~J(\{\response_\time\}),\text{ where}\label{eqn:noise_irp}\\ J(\{\response_\time\})=& \sum_{\time=1}^\horizon (u(\response_\time)-u(\idresponse_\time)) - \lambda~\prob(H_1|\{\probe_\time,\response_\time,u\}).\nonumber
\end{align}
In \eqref{eqn:noise_irp}, 
$\lambda>0$ is a pre-specified scalar that parametrizes the degree of meta-cognition of the radar and $\prob(H_1|\{\probe_\time,\response_\time,u\})$ is the conditional Type-I error probability of the adversary's detector~\eqref{eqn:stat_test} defined in \eqref{eqn:surr_error}.
\end{theorem}

\begin{algorithm} \caption{Meta-cognition. SPSA-based  Stochastic Gradient Algorithm for Confusing Adversary's IRL Detector (Theorem~\ref{thrm:noise_irp})} \label{alg:noise_irp}
Given a sequence of adversary's probe signals $\{\probe_\time\}_{\time=1}^{\horizon}$, output  radar's I-IRL responses $\vecresponse=\{\pertresponse_\time\}_{\time=1}^{\horizon}$ \eqref{eqn:noise_irp} that confuses the adversary's detector \eqref{eqn:stat_test}.
\hspace{-0.5cm}\begin{compactitem}
\item[Step 1.] Choose initial response $\vecresponse_0=\{\response_{0,\time}\}_{\time=1}^{\horizon}$ as the naive response sequence:~$\vecresponse_0=\{\arg\max_{\{\response:\probe_\time'\response\leq 1\}}u(\response)\}_{\time=1}^{\horizon}$.
\item[Step 2.] For iterations $\iter=0,1,2,\ldots$~:\\ \vspace{0.2cm}
(i) Estimate cost $J(\vecresponse_\iter)$, $\vecresponse_\iter=\{\response_{\iter,\time}\}$ in \eqref{eqn:noise_irp} as:
    \begin{align}
\hspace{-0.65cm}  &\hat{J}(\vecresponse_\iter)  = \sum_{\time=1}^\horizon(u(\response_{\iter,\time})-u(\response_{0,\time})) - \lambda~\empprob(H_1|\{\probe_\time\},\vecresponse_\iter,u), \nonumber\\
   \hspace{-0.65cm}& \empprob(H_1|\{\probe_\time\},\vecresponse_\iter,u)=\frac{1}{R}\sum_{r=1}^{R}\mathbbm{1}\{F_L(\phi^\ast_u(\{\probe_\time\},\widehat{\vecresponse}_{\iter,r}))\geq 1-\alpha\},\nonumber\\
   \hspace{-0.65cm}& \widehat{\vecresponse}_{\iter,r} = \{\response_{\iter,\time} + w_{\time,r}\}_{\time=1}^{\horizon},~w_{\time,r}\sim f_w. \label{eqn:emp_objcompute}
\end{align}
In \eqref{eqn:emp_objcompute}, $\mathbbm{1}\{\cdot\}$ denotes the indicator function, $F_L(\cdot)$ is the distribution function of the random variable $L$ defined in \eqref{eqn:stat_test}. The sufficient statistic $\phi^\ast_u(\cdot)$ is defined in \eqref{eqn:surr_error}.  The variable $\widehat{\vecresponse}_\iter=\vecresponse_\iter+\{w_{\time,r}\}$ is a noisy observation of the radar's response, and $w_{\time,r}$ is a fixed realization of random variable  $w_\time$~\eqref{eqn:stat_test}. The parameter $R$ controls the accuracy of the empirical probability estimate, and parameter $\alpha$ is the significance level of the adversary's statistical hypothesis test \eqref{eqn:stat_test} that upper bounds its Type-I error probability.\vspace{0.15cm}\\
(ii) Estimate the gradient $\nabla_{\vecresponse}J(\vecresponse_\iter)$ as:
\begin{equation}\label{eqn:compute_empgrad}
    \widehat{\nabla}_{\vecresponse}(\hat{J}(\vecresponse_\iter)) = \frac{\Delta_\iter}{2\omega\|\Delta_\iter\|_F^2}~\hat{J}(\vecresponse_\iter + \omega~\Delta_\iter) - \hat{J}(\vecresponse_\iter - \omega~\Delta_\iter),
\end{equation}
where $\omega$ is the gradient step size, $\hat{J}(\cdot)$ is defined in \eqref{eqn:emp_objcompute}, and $\Delta_\iter\in\reals^{\probedim\times\horizon}$ is a random perturbation vector whose each element is set to $+1$ or $-1$ with probability $1/2$.\\ \vspace{0.2cm}
(iii) Update the radar's response as:
\begin{equation}\label{eqn:SPSA_update}
 \vecresponse_{\iter+1} = \operatorname{Proj}_{S}\left(\vecresponse_\iter - \eta~\frac{\Delta_\iter}{\|\Delta_\iter\|_F}\widehat{\nabla}_{\vecresponse} \hat{J}(\vecresponse_\iter)\right),
\end{equation}
where $\operatorname{Proj}_{S}$ is the projection operator to the hyperplane $S = \{\vecresponse: \probe_\time'\response_\time=1,\response_\time\geq 0\}$ and $\eta>0$ is the step size.
\item[Step 3.] Set $\iter\leftarrow \iter+1$ and go to Step 2.
\end{compactitem}
\end{algorithm}

The meta-cognitive radar in Theorem~\ref{thrm:noise_irp} trades off between two quantities: (1) {\em Maximizing quality of service (QoS)} by minimizing the utility loss due to the perturbed 
I-IRL response, and (2) {\em Maximizing the adversary's confusion} by maximizing the conditional Type-I error probability~\eqref{eqn:surr_error} of the adversary's statistical test \eqref{eqn:stat_test} in Theorem~\ref{thrm:noise_rp}.

The optimization problem in \eqref{eqn:noise_irp} can be solved by a stochastic gradient algorithm, such as SPSA~\cite{Spa03}, outlined in Algorithm~\ref{alg:noise_irp}. SPSA is a generalization of adaptive algorithms where the gradient computation in \eqref{eqn:noise_irp} requires only two measurements of the objective function corrupted by noise per iteration, i.e.\,, the number of evaluations is independent of the dimension of $\{\response_\time\}$; see \cite{Spa03} for a tutorial exposition of the SPSA algorithm. For decreasing step size $\eta=1/i$ \eqref{eqn:SPSA_update}, the SPSA algorithm converges with probability one to a local stationary point. For constant step size $\eta$, it converges weakly (in probability).

{\em Degree of meta-cognition $\lambda$ for Theorem~\ref{thrm:noise_irp}.} A larger value of $\lambda$~\eqref{eqn:noise_irp} results in a larger Type-I error probability for the adversary's IRL detector, while increasing the radar's deviation from its optimal response. A few words on how the I-IRL approach for Theorem~\ref{thrm:noise_irp} differs from Theorem~\ref{thrm:irp}. The key idea in Theorem~\ref{thrm:irp} is to push the true utility function away from the center of the feasible set of utilities reconstructed by the adversary. In comparison, Theorem~\ref{thrm:noise_irp} maximizes the conditional Type-I error probability~\eqref{eqn:surr_error} of the adversary's detector \eqref{eqn:stat_test}. The variable $\lambda$ parametrizes the extent of meta-cognition Theorem~\ref{thrm:noise_irp}, in complete analogy to the parameter $\epsilon$ in Theorem~\ref{thrm:irp}.

{\em Summary.} This section presented two I-IRL results for meta-cognitive radars as counter-adversarial measures to obfuscate IRL algorithms in Sec.\,\ref{sec:background} used by an adversary to estimate the radar's cognitive ability (utility function). The first I-IRL result, Theorem~\ref{thrm:irp}, achieves meta-cognition in the noise-less setting where the adversary has accurate measurements of the radar's responses and uses Theorem~\ref{thrm:rp} for IRL. The second I-IRL result, Theorem~\ref{thrm:noise_irp}, achieves meta-cognition when the adversary measures the radar's response in noise and uses a utility maximization detector (Theorem~\ref{thrm:noise_rp}) for IRL. Although counter-intuitive, achieving meta-cognition in the noisy case is more difficult compared to that in the noise-less case due to the increased robustness of the IRL algorithm (statistical hypothesis test vs.\ a deterministic feasibility test) used by the adversary. In the next section, we illustrate the I-IRL results of Theorems~\ref{thrm:irp} and \ref{thrm:noise_irp} through simple numerical examples.

\section{Numerical Results}\label{sec:numericalresults}
In this section, we present two numerical examples to illustrate the how the I-IRL results of Sec.\,\ref{sec:iirl} successfully mitigate the IRL algorithms (presented in Sec.\,\ref{sec:background}) used by the adversarial target.
\subsection{Numerical Example 1. I-IRL for Afriat's Theorem (Theorem~\ref{thrm:rp}) using Theorem~\ref{thrm:irp}}
We chose $\horizon=50$ and  $\sigdim=2$, the dimension of adversarial target's probe and radar's response. Recall from Sec.\,\ref{sec:radar_example} that the probe signal parameterizes the state covariance matrix of the radar's Kalman filter due to the adversary's maneuvers, and the response signal parameterizes the sensory accuracy chosen by the radar. The elements of the adversarial target's probe signals are generated randomly and independently over time as $\probe_k(i)\sim$ Unif$(0.2,2.5)$ for all $i=1,2$ and time $\time=1,2,\ldots,\horizon$,where Unif($a,b$) denotes uniform pdf with support $(a, b)$. Recall  that the probe signal $\probe_\time$ is the diagonal of the state noise covariance matrix: $\statenoisecov_\time=\operatorname{diag}[\probe_\time(1),\probe_\time(2)]$.

\begin{figure}[h]
    \centering
  \includegraphics[width=0.85\linewidth]{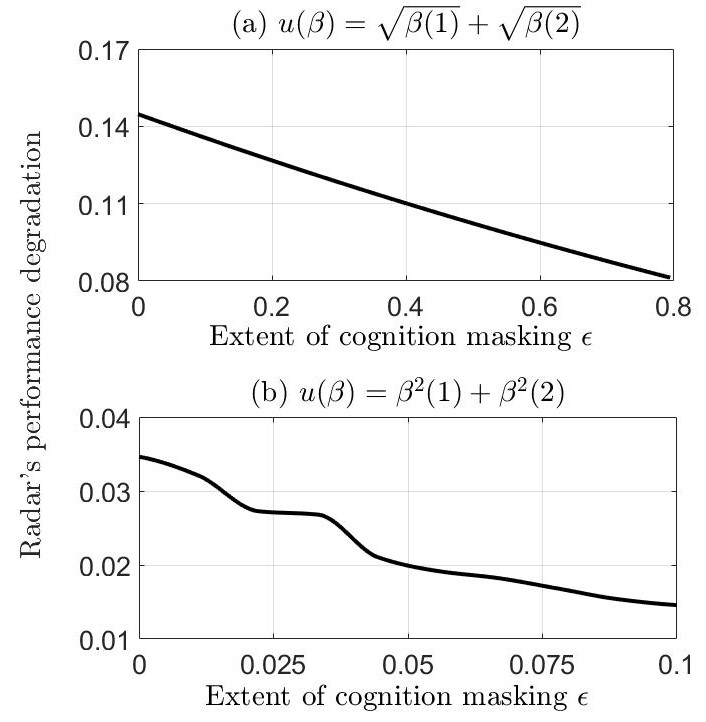}
    \caption{The I-IRL approach  in this paper: Small deliberate performance loss (vertical axis) of the meta-cognitive radar results in large performance loss of the IRL adversary (horizontal axis). The figure illustrates a meta-cognitive radar operating with two distinct utility functions.
    (i) $\epsilon=0$ corresponds to maximum cognition masking and hence results in maximum performance loss. (ii) Due to larger local variation, for a fixed value of $\epsilon$, the quadratic utility (sub-figure (b)) requires smaller perturbation ($\approx$ 10 times) from the optimal response 
    compared to the linear utility (sub-figure (a)). }
    \label{fig:detconfplot}
\end{figure}

Given the probe sequence $\{\probe_\time,\time=1,2,\ldots,\horizon\}$, the cognitive radar chooses its response sequence $\{\response_\time,\time=1,2,\ldots,\horizon\}$ via \eqref{eqn:irp} in Theorem~\ref{thrm:irp}. Recall from Sec.\,\ref{sec:background} that response $\response_\time$ is the diagonal of the inverse of radar's observation noise covariance matrix: $\obsnoisecov_\time^{-1} = \operatorname{diag}[\response_\time(1),\response_\time(2)]$ .
We generate two separate sequences of responses for the same probe sequence, but for two different utility functions:
\begin{align*}
    (a)&~u(\response) = \sqrt{\response(1)}  + \sqrt{\response(2)},\qquad (b)~u(\response) = \response^2(1) + \response^2(2)
\end{align*}
Figure~\ref{fig:detconfplot} shows the loss in performance (minimum perturbation from optimal response~\eqref{eqn:irp}) of the cognitive radar as a function of $\epsilon$ (extent of cognition masking), for both choices of utility functions.
%
From Fig.\,\ref{fig:detconfplot}, we see that for both utility functions, the radar's performance decreases with increasing $\epsilon$ (larger extent of utility masking). This is expected since larger $\epsilon$ implies larger shift of the feasible set of utilities constructed by the adversarial target.

\subsection{Numerical Example 2. I-IRL for stochastic Afriat's Theorem (Theorem~\ref{thrm:noise_rp}) using Theorem~\ref{thrm:noise_irp}}
For the second numerical example, we chose $\horizon=20$ and  $\sigdim=3$. The elements of the adversarial target's probe signals are generated randomly and independently over time as $\probe_k(i)\sim$ Unif$(1,4)$ for all $i=1,2,3$ and time $\time=1,2,\ldots,\horizon$. Recall that the adversarial target observes the radar's responses in noise. We set the noise pdf $f_w$~\eqref{eqn:noise_response} to $\mathcal{N}(0,0.2~I)$, where $\mathcal{N}(\mu,\Sigma)$ denotes the multivariate normal distribution with mean $\mu$ and covariance $\Sigma$, and $I$ denotes the identity matrix. In Theorem~\ref{thrm:noise_irp},
We chose the radar's utility function $u$ to be $u(\response)=\sum_{i=1}^3\sqrt{\response(i)}$ and performed our numerical experiment for three values of $\alpha=\{0.05,0.1,0.2\}$, the significance level $\alpha$~\eqref{eqn:stat_test} in the adversary's statistical test.

Given the probe sequence $\{\probe_\time,\time=1,2,\ldots,\horizon\}$, we generated the meta-cognitive radar's response sequence via Theorem~\ref{thrm:noise_irp} by varying the parameter $\lambda$~\eqref{eqn:noise_irp} over the interval $[10^0,10^5]$. Our SPSA algorithm in Algorithm~\ref{alg:noise_irp} was executed over $10^5$ iterations for each value of $\lambda$ and $\alpha$. Table \ref{tab:noise_irp} shows the conditional Type-I error probability (adversary's confusion) of the adversary's detector and utility loss of the radar due to its sub-optimal response as the parameter $\lambda$ is varied for three different values of the significance level $\alpha$ of the adversary's detector. From Table \ref{tab:noise_irp}, we see that both the radars' utility loss and adversary's confusion increase with both parameters $\lambda$ and $\alpha$. It is straightforward to justify the variation of the adversary's confusion and radar's utility loss due to the parameter $\lambda$ that scales the error probability term in the radar's meta-cognition objective function, $J$~\eqref{eqn:noise_irp}. If $\lambda=0$, the I-IRL response of the radar would be simply the optimal waveform that maximizes the radar's utility, resulting in zero utility loss for the radar. For the limiting case of $\lambda\to\infty$, the radar's I-IRL response computed via Theorem~\ref{thrm:noise_irp} degenerates to a constant for all time $\time$, hence maximizing the conditional Type-I error probability of the detector at the cost of maximal loss of utility.

The variation of the adversary's confusion with the significance level of the adversary's detector is a key feature that warrants more discussion. The parameter $\alpha$~\eqref{eqn:stat_test} can be viewed as the risk-aversion of the adversarial target. A larger value of $\alpha$ implies a greater bound on the Type-I error probability of the adversary's detector. That is, under the null hypothesis the radar is a utility maximizer, the set of noisy datasets $\ndataset$ for which the adversary's detector classifies the radar as not a utility maximizer gets larger with $\alpha$. In other words, a smaller perturbation (equivalently, smaller utility loss) in the radar's response suffices to reject the null hypothesis and thus results in a larger conditional Type-I error probability for the adversary's detector.

\begin{figure}[h]
    \centering
    {\small Adversary IRL detector's conditional Type-I error\\ probability due to I-IRL (Theorem~\ref{thrm:noise_irp})\vspace{0.15cm}\\
    \begin{tabular}{|c|c|c|c|}\hline
        $(\lambda,\alpha)$ & $\alpha=0.05$ & $\alpha=0.1$ & $\alpha=0.2$ \\ \hline
        $\lambda=10^0$ & 0.11 & 0.19 & 0.36\\\hline
        $\lambda=10^1$ & 0.17 & 0.34 & 0.49\\\hline
        $\lambda=10^2$ & 0.25 & 0.36 & 0.61\\\hline
        $\lambda=10^3$ & 0.68 & 0.79 & 0.79\\\hline
        $\lambda=10^4$ & 0.74 & 0.8 & 0.83 \\\hline
        $\lambda=10^5$ & 0.93 & 1 & 1 \\\hline
    \end{tabular}\vspace{0.3cm}\\
     Radar's performance loss due to I-IRL (Theorem~\ref{thrm:noise_irp})\vspace{0.15cm}\\
    \begin{tabular}{|c|c|c|c|}\hline
        $(\lambda,\alpha)$ & $\alpha=0.05$ & $\alpha=0.1$ & $\alpha=0.2$ \\ \hline
        $\lambda=10^0$ & $0.001$ & $0.004$ & $0.02$\\\hline
        $\lambda=10^1$ & $0.008$ & $0.01$ & $0.043$\\\hline
        $\lambda=10^2$ & $0.005$  & $0.03$ & $0.06$\\\hline
        $\lambda=10^3$ & $0.043$ & $0.06$ & $0.06$\\\hline
        $\lambda=10^4$ & $0.077$ & $0.095$ & $0.1$\\\hline
        $\lambda=10^5$ & $0.05$ & $0.3$ & 0.4  \\\hline
    \end{tabular}}
    \caption{ I-IRL performance of meta-cognitive Radar when the adversary
    has noisy measurements of the radar's response and deploys an IRL detector. The key takeaway is that a  small performance loss of the radar results results in large performance loss of adversary's IRL detector. The performance loss of both the radar and the adversary due to meta-cognition increase with scaling factor $\lambda$~\eqref{eqn:noise_irp} and significance level $\alpha$ of the adversary's IRL detector \eqref{eqn:stat_test}.}
    \label{tab:noise_irp}
\end{figure}

\section{Conclusion and Extensions}
This paper studies the interaction  between a cognitive radar and a smart adversary when the cognitive radar is aware of the smart adversary.
We develop an inverse inverse reinforcement learning (I-IRL) based approach to design a meta-cognitive radar to  mask the  radar's utility function when probed by an adversarial target. The main takeaway is that the meta-cognitive radar causes a large disturbance in the adversary's IRL algorithm at the cost of small performance loss in its utility. This is in contrast to a traditional cognitive radar that maximizes its performance but is at maximum risk of revealing its utility function to the adversary.

Our main I-IRL results are Theorems~\ref{thrm:irp} and \ref{thrm:noise_irp}. These   specify the meta-cognitive radar's strategy when the adversarial target uses IRL to estimate the radar's utility function. Theorem~\ref{thrm:irp} achieves I-IRL when the adversary measures the radar's responses accurately and uses a deterministic convex feasibility test to estimate the radar's utility. Theorem~\ref{thrm:noise_irp} assumes a more sophisticated adversary that measures the radar's response in noise and
uses a statistical hypothesis test to detect the radar's utility function.
The key idea behind both meta-cognition results is to sufficiently confuse the adversary by deliberately choosing  sub-optimal responses at the cost of the radar's performance. In Theorem~\ref{thrm:irp}, the sub-optimal response of the radar ensures the radar's true utility function passes the adversary's feasibility test by a low margin. In Theorem~\ref{thrm:noise_irp}, the radar's response increases the conditional Type-I error probability of the adversary's detector and tricks the adversary's detector into classifying the radar as non-cognitive with high probability.

This paper has focused on the radar hiding its utility function from the adversary.
The methods can be easily extended to the design
of  meta-cognitive radars that hide their resource constraints instead of their utility functions. Such situations arise in scenarios where the radar wants to hide its constraint capability from a smart adversary. Finally, a natural extension is to formulate the radar-adversary interaction as a game, where now the adversary aims to mitigate the radar's I-IRL scheme, and identify play from their Nash equilibrium.

\bibliographystyle{unsrt_abbrv_custom}
\bibliography{refs}

\begin{thebibliography}{10}

\bibitem{SUT18}
R.~S. Sutton and A.~G. Barto.
\newblock {\em Reinforcement learning: An introduction}.
\newblock MIT press, 2018.

\bibitem{KA18}
L.~Kang, J.~Bo, L.~Hongwei, and L.~Siyuan.
\newblock Reinforcement learning based anti-jamming frequency hopping
  strategies design for cognitive radar.
\newblock In {\em 2018 IEEE International Conference on Signal Processing,
  Communications and Computing (ICSPCC)}, pages 1--5. IEEE, 2018.

\bibitem{ABB01}
J.~Abounadi, D.~P. Bertsekas, and V.~Borkar.
\newblock Learning algorithms for {M}arkov decision processes with average
  cost.
\newblock {\em SIAM Journal on Control and Optimization}, 40(3):681--698, 2001.

\bibitem{Afr67}
S.~Afriat.
\newblock The construction of utility functions from expenditure data.
\newblock {\em International economic review}, 8(1):67--77, 1967.

\bibitem{Var83}
H.~Varian.
\newblock Non-parametric tests of consumer behaviour.
\newblock {\em The Review of Economic Studies}, 50(1):99--110, 1983.

\bibitem{Hay06}
S.~Haykin.
\newblock Cognitive radar.
\newblock {\em IEEE Signal Processing Magazine}, pages 30--40, Jan. 2006.

\bibitem{Hay12}
S.~Haykin.
\newblock Cognitive dynamic systems: Radar, control, and radio [point of view].
\newblock {\em Proceedings of the IEEE}, 100(7):2095--2103, 2012.

\bibitem{BBSJR15}
K.~Bell, C.~Baker, G.~Smith, J.~Johnson, and M.~Rangaswamy.
\newblock Cognitive radar framework for target detection and tracking.
\newblock {\em IEEE Journal of Selected Topics in Signal Processing},
  9(8):1427--1439, 2015.

\bibitem{KAEM20}
V.~Krishnamurthy, D.~Angley, R.~Evans, and B.~Moran.
\newblock Identifying cognitive radars - inverse reinforcement learning using
  revealed preferences.
\newblock {\em IEEE Transactions on Signal Processing}, 68:4529--4542, 2020.

\bibitem{KPGKR21}
V.~Krishnamurthy, K.~Pattanayak, S.~Gogineni, B.~Kang, and M.~Rangaswamy.
\newblock Adversarial radar inference: Inverse tracking, identifying cognition,
  and designing smart interference.
\newblock {\em IEEE Transactions on Aerospace and Electronic Systems},
  57(4):2067--2081, 2021.

\bibitem{DFIG1}
E.~Blasch, I.~Kadar, J.~Salerno, M.~M. Kokar, S.~Das, G.~M. Powell, D.~D.
  Corkill, and E.~H. Ruspini.
\newblock Issues and challenges of knowledge representation and reasoning
  methods in situation assessment (level 2 fusion).
\newblock In {\em Signal Processing, Sensor Fusion, and Target Recognition XV},
  volume 6235, page 623510. International Society for Optics and Photonics,
  2006.

\bibitem{KR19}
V.~Krishnamurthy and M.~Rangaswamy.
\newblock How to calibrate your adversary's capabilities? inverse filtering for
  counter-autonomous systems.
\newblock {\em IEEE Transactions on Signal Processing}, 67(24):6511--6525,
  2019.

\bibitem{PK20inverse}
K.~Pattanayak, V.~Krishnamurthy, and E.~Blasch.
\newblock Inverse sequential hypothesis testing.
\newblock In {\em 2020 IEEE 23rd International Conference on Information Fusion
  (FUSION)}, pages 1--7. IEEE, 2020.

\bibitem{BD78}
J.~A. Boyd, D.~B. Harris, D.~D. King, and H.~Welch~Jr.
\newblock Electronic countermeasures.
\newblock {\em Electronic Countermeasures}, 1978.

\bibitem{SC86}
D.~C. Schleher.
\newblock Introduction to electronic warfare.
\newblock {\em Dedham}, 1986.

\bibitem{PKB21}
K.~Pattanayak, V.~Krishnamurthy, and C.~Berry.
\newblock How can a cognitive radar mask its cognition?
\newblock {\em arXiv preprint arXiv:2110.08608}, 2021.

\bibitem{ECCMsurvey}
L.~Neng-Jing and Z.~Yi-Ting.
\newblock A survey of radar ecm and eccm.
\newblock {\em IEEE Transactions on Aerospace and Electronic Systems},
  31(3):1110--1120, 1995.

\bibitem{ECCM1}
C.~Shi, F.~Wang, M.~Sellathurai, and J.~Zhou.
\newblock Low probability of intercept-based distributed mimo radar waveform
  design against barrage jamming in signal-dependent clutter and coloured
  noise.
\newblock {\em IET Signal Processing}, 13(4):415--423, 2019.

\bibitem{ECCM2}
F.~A. Butt, I.~H. Naqvi, and U.~Riaz.
\newblock Hybrid phased-mimo radar: A novel approach with optimal performance
  under electronic countermeasures.
\newblock {\em IEEE Communications Letters}, 22(6):1184--1187, 2018.

\bibitem{stealth1}
W.-Q. Wang.
\newblock Moving-target tracking by cognitive rf stealth radar using frequency
  diverse array antenna.
\newblock {\em IEEE Transactions on Geoscience and Remote Sensing},
  54(7):3764--3773, 2016.

\bibitem{stealth2}
W.-Q. Wang.
\newblock Adaptive rf stealth beamforming for frequency diverse array radar.
\newblock In {\em 2015 23rd European Signal Processing Conference (EUSIPCO)},
  pages 1158--1161. IEEE, 2015.

\bibitem{stealth3}
Z.~Zhang, S.~Salous, H.~Li, and Y.~Tian.
\newblock Optimal coordination method of opportunistic array radars for
  multi-target-tracking-based radio frequency stealth in clutter.
\newblock {\em Radio Science}, 50(11):1187--1196, 2015.

\bibitem{NG00}
A.~Y. Ng, S.~J. Russell, et~al.
\newblock Algorithms for inverse reinforcement learning.
\newblock In {\em Icml}, volume~1, page~2, 2000.

\bibitem{ZB08}
B.~D. Ziebart, A.~L. Maas, J.~A. Bagnell, A.~K. Dey, et~al.
\newblock Maximum entropy inverse reinforcement learning.
\newblock In {\em Aaai}, volume~8, pages 1433--1438. Chicago, IL, USA, 2008.

\bibitem{Afr87}
S.~Afriat.
\newblock {\em Logic of choice and economic theory}.
\newblock Clarendon Press Oxford, 1987.

\bibitem{Die12}
W.~Diewert.
\newblock Afriat's theorem and some extensions to choice under uncertainty.
\newblock {\em The Economic Journal}, 122(560):305--331, 2012.

\bibitem{Var12}
H.~Varian.
\newblock Revealed preference and its applications.
\newblock {\em The Economic Journal}, 122(560):332--338, 2012.

\bibitem{Var06}
H.~Varian.
\newblock Revealed preference.
\newblock {\em Samuelsonian economics and the twenty-first century}, pages
  99--115, 2006.

\bibitem{Var82}
H.~Varian.
\newblock The nonparametric approach to demand analysis.
\newblock {\em Econometrica}, 50(1):945--973, 1982.

\bibitem{KH12}
V.~Krishnamurthy and W.~Hoiles.
\newblock Afriat's test for detecting malicious agents.
\newblock {\em IEEE Signal Processing Letters}, 19(12):801--804, 2012.

\bibitem{LJ03}
X.~R. Li and V.~P. Jilkov.
\newblock Survey of maneuvering target tracking. part i. dynamic models.
\newblock {\em IEEE Transactions on Aerospace and Electronic Systems},
  39(4):1333--1364, 2003.

\bibitem{BLK08}
Y.~Bar-Shalom, X.~R. Li, and T.~Kirubarajan.
\newblock {\em Estimation with applications to tracking and navigation}.
\newblock John Wiley, New York, 2008.

\bibitem{BP99}
S.~Blackman and R.~Popoli.
\newblock {\em Design and Analysis of Modern Tracking Systems}.
\newblock Artech House, 1999.

\bibitem{Spa03}
J.~Spall.
\newblock {\em Introduction to Stochastic Search and Optimization}.
\newblock Wiley, 2003.

\end{thebibliography}

\end{document}